\documentclass[twocolumn,showpacs,preprintnumbers,amsmath,amssymb,10pt,prd]{revtex4}

\usepackage{graphicx}% Include figure files

\usepackage{dcolumn}% Align table columns on decimal point

\usepackage{bm}% bold math

\usepackage{hyperref}

\usepackage{mathrsfs}

\def\Hc{\mathscr{H}}

\newcommand{\lb}[1]{\label{#1}}

\newcommand{\bark}{\bar{k}}

\newcommand{\barp}{\bar{p}}

\newcommand{\barmu}{\bar{\mu}}

\newcommand{\barN}{\bar{N}}

\newcommand\be{\begin{equation}}

\newcommand\ee{\end{equation}}

\newcommand\ba{\begin{eqnarray}}

\newcommand\ea{\end{eqnarray}}

\begin{document}

\title{Primordial tensor power spectrum in holonomy corrected $\mathbf{\Omega}$ loop
quantum cosmology}

%\author{Sex, Truck \& Rock'n Roll}

\author{Linda Linsefors}%

 \email{linsefors@lpsc.in2p3.fr}

\affiliation{%
Laboratoire de Physique Subatomique et de Cosmologie, UJF, INPG, CNRS, IN2P3\\
53, avenue des Martyrs, 38026 Grenoble cedex, France
}%

\author{Thomas Cailleteau}%

 \email{thomas@gravity.psu.edu}

\affiliation{%
Institute for Gravitation \& the Cosmos, Penn State\\
University Park, PA 16802, USA
}

\author{Aurelien Barrau}%

 \email{Aurelien.Barrau@cern.ch}

\affiliation{%
Laboratoire de Physique Subatomique et de Cosmologie, UJF, INPG, CNRS, IN2P3\\
53,avenue des Martyrs, 38026 Grenoble cedex, France
}%

\author{Julien Grain}

 \email{julien.grain@ias.u-psud.fr}

 \affiliation{%
 Universit\'e Paris-Sud 11, Institut d'Astrophysique Spatiale, UMR8617, 91405 Orsay, France}
\affiliation{%
CNRS, Orsay, France, F-91405}

\date{\today}

\begin{abstract}

The holonomy correction is one of the main terms arising when implementing loop quantum
gravity ideas at an effective level in cosmology. The recent construction of an anomaly
free algebra has shown that the formalism used, up to now, to derive the primordial spectrum of
fluctuations was not correct. This article aims at computing the tensor spectrum in a fully
consistent way within this deformed and closed algebra.

\end{abstract}


% PACS, the Physics and Astronomy

% Classification Scheme.

%Use showkeys class option if keyword

\maketitle

\section{Introduction}

Nonperturbatively quantizing General Relativity 
(GR) in a background-invariant way is obviously an outstanding open problem of theoretical physics. Loop
Quantum Gravity (LQG) is a promising framework in which to perform this program (see \cite{lqg_review} for introductory 
reviews). Although this is still to be demonstrated, there is evidence that different
approaches, based either on  quantizations (covariant or canonical) of GR, or on 
a formal quantization of geometry, lead to
the same LQG theory. Experimental tests are, however, still missing.
Trying to find possible observational signatures is  a key
challenge and cosmological footprints are known for being one of the only possible paths toward a real experimental
test of LQG. It is very hard to make clear predictions in Loop Quantum Cosmology (LQC) using the
full ``mother" LQG theory. General introductions to LQC can be found in  \cite{lqc_review}. This study focuses on
an effective treatment taking into account recent results on the correct algebra of constraints. 
We first review the theoretical framework. The spectrum is then derived. Some
conclusions and consequences are finally underlined.

\section{Theoretical Framework}

One of the fundamental quantum corrections expected from the Hamiltonian of LQG arises from the fact that loop 
quantization is based on holonomies, {\it i.e.}, exponentials of the connection, rather than direct connection 
components. Based on a canonical approach, the theory uses Ashtekar variables, namely, SU(2) valued connections and 
conjugate densitized triads. The quantization is obtained through holonomies of the connections and fluxes of the 
densitized triads.

This is the key ingredient of the effective approach. The cosmological equations are modified so as to account for 
the loop basis of the theory.\\

The main consequence of the holonomy correction on the cosmological background is to induce a bounce. The evolution 
is not singular anymore and the Big Bang
is replaced by a Big Bounce. The next step consists of studying the propagation of perturbations within this modified 
background.

In cosmology, perturbations are of three different types : scalar, vector, and tensor. We focus here on the tensor 
modes that are directly gauge-invariant.
Quite a lot of works have already been devoted to tensor modes in this framework \cite{pheno_tensor}. Beyond
that, 
the phenomenology of LQG is now a well-established field (see \cite{pheno} for a review). Unfortunately, a recent 
study \cite{tom} has shown that the previously derived spectra are most probably incorrect.\\

The key issue relies in the closure of the algebra of constraints. Due to general covariance, the
canonical Hamiltonian is a combination of constraints  $\mathcal{C}_I$. Consistency requires
that the constraints are preserved under the evolution they generate. This is ensured in the classical theory by the 
closure of the Poisson algebra of constraints

\begin{equation}
\{ \mathcal{C}_I, \mathcal{C}_J \} = {f^K}_{IJ}(A^j_b,E^a_i) \mathcal{C}_K, \label{algebra}
\end{equation}

where $\mathcal{C}_I$, $I=1,2,3,$ are the Gauss, diffeomorphism and Hamiltonian constraints and
${f^K}_{\!IJ}( A^j_b,E^a_i)$ are structure functions which, in general,
depend on the phase space (Ashtekar) variables  $(A^j_b, E^a_i)$. They form a first class set. Otherwise stated, 
the gauge transformations and evolution generated by the constraints define vector fields that are tangent to the 
submanifold defined by the vanishing of constraints.

In LQC, quantum corrections are introduced as effective modifications of the
Hamiltonian constraint. This generates anomalies: the modified
constraints $\mathcal{C}^Q_I$ do not
form  a closed algebra anymore,

\begin{equation}
\{ \mathcal{C}^Q_I, \mathcal{C}^Q_J \} = {f^K}_{IJ}(A^j_b,E^a_i)\, \mathcal{C}^Q_K+
\mathcal{A}_{IJ}.
\end{equation}

The anomalous terms $\mathcal{A}_{IJ}$ are removed by carefully adjusting the form of the quantum correction to the 
Hamiltonian constraint through the addition of suitable ``counterterms" that vanish in the classical limit. This has
 been done in \cite{tom}, following the approach of \cite{anomaly}.

In the classical case, the Poisson brackets between the constraints read as
\\[-5mm]

\ba \lb{algtot} &\hspace{-9mm}
\{D_{(\!m\!+\!g\!)}[N^a_1],D_{(\!m\!+\!g\!)}[N^a_2]\} = 0 \,,
\hspace{-5mm} &\\ &\hspace{-9mm}
\{H_{(\!m\!+\!g\!)}[N],D_{(\!m\!+\!g\!)}[N^a]\} = - H_{(\!m\!+\!g\!)}[\delta N^a \partial_a \delta N] \,,
\hspace{-5mm}& \\ &\hspace{-6mm}
\{H_{(\!m\!+\!g\!)}[N_1], H_{(\!m\!+\!g\!)}[N_2]\} = D_{(\!m\!+\!g\!)}\!\!\left[\!\frac{\barN}{\barp} \partial^a (\delta N_2 - \delta N_1) \!\right]\!,~
\lb{Dmg}
\ea

where $(\!m\!+\!g\!)$ stands for gravity and matter. The quantum corrections are included at the effective level by 
replacing, as usual, in the Hamiltonian constraint

\be
\bark \to  \frac{\sin ( \barmu \gamma \bark)}{ \barmu \gamma}.
\ee

The important result of \cite{tom} is that the quantum-corrected algebra is described by a single modification:

\ba
\{H_{(\!m\!+\!g\!)}[N_1], H_{(\!m\!+\!g\!)}[N_2]\} = {\bf \Omega} \; D_{(\!m\!+\!g\!)}\!\!\left[\frac{\barN}{\barp} \partial^a (\delta N_2 - \delta N_1) \right]%\!,
\nonumber\hspace{-5mm}&\\ \lb{Dmq}
\ea\\[-7mm]

where\\[-7mm]

\begin{equation}
{\bf \Omega} = \cos(2\barmu \gamma \bark) = 1- 2 \frac{\rho}{\rho_c}~.
\label{omega}
\end{equation}

The  ${\bf \Omega}$ factor encodes the quantum correction, $\bark$ being the homogeneous Ashtekar connection and 
$\barmu$ being proportional to the ratio between the Planck length and the scale factor.
The Mukhanov-Sasaki \cite{Mukhanov:1990me} equation of motion for gauge-invariant perturbations of scalar and tensor 
types $v_{\mathrm{S(T)}}$ can be explicitly derived. In conformal time, the propagation of tensor modes is given by 

\begin{equation}\lb{EoM}
%\ddot{v} - \Omega \, \nabla ^2 v - \frac{\ddot{z}}{z} v= 0,
{v}''_{\mathrm{T}} - {\bf \Omega} \, \nabla ^2 v_{\mathrm{T}} - \frac{{z''_{\mathrm{T}}}}{z_{\mathrm{T}}} v_{\mathrm{T}}= 0
~;~z_{\mathrm{T}}=\frac{a}{\sqrt{\mathbf{\Omega}}},
\end{equation}

where \textit{prime} means differentiation with respect to conformal time.
This leads to the following equation of motion for tensor perturbations, defined via 
$v_\mathrm{T}=z_\mathrm{T}\times h^i_a$:

\begin{equation}
{h^i_a}'' + {h^i_a}'
 \; \left(2\Hc - {\frac{{\bf \Omega}'}{{\bf \Omega}}}\right) - {\bf \Omega} \, \nabla ^2 {h^i_a} = 0,
\label{eqh}
\end{equation}

where $\Hc:=a'/a$ is the conformal Hubble parameter.

\section{Power spectrum}

This equation being known, it is possible to investigate the associated primordial power spectrum. This is the 
fundamental ingredient for phenomenology. The background dynamics is not modified by the $\mathbf{\Omega}$ term. 
However, the perturbations will, of course, undergo a different evolution.

\begin{figure}

	\begin{center}
		\includegraphics[scale=1]{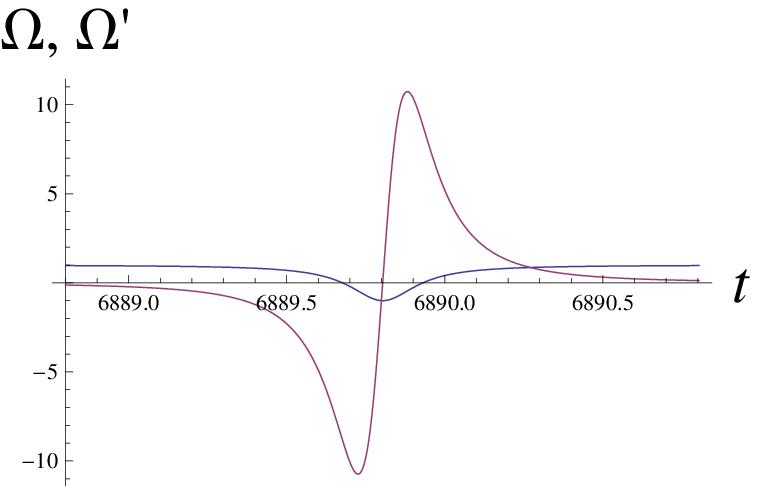}
		\caption{Evolution of $\mathbf{\Omega}$ and its derivative with respect to conformal time.			
		The density where $\mathbf{\Omega}$ vanishes is half the critical density
		whereas $\mathbf{\Omega}'$ vanishes at the bounce.}
		\label{fig1}
	\end{center}
\end{figure}

We use the Fourier transformed version of Eq. (\ref{eqh}):

\begin{equation}
{h}''+\left(2\Hc-\frac{{\mathbf{\Omega}}'}{\mathbf{\Omega}}\right){h}'+\mathbf{\Omega} k^2h=0,
\label{eomk}
\end{equation}

where the indices have been skipped for simplicity. The behavior of $\mathbf{\Omega}$ and 
${\mathbf{\Omega}}'$ is displayed in Fig. \ref{fig1}. One can immediately see that $\mathbf{\Omega}$ vanishes for 
$\rho=\rho_c/2$, where

\begin{equation}
\rho_{\text{c}} = \frac{\sqrt{3}}{32\pi^2\gamma^3}m^4_{\text{Pl}} \simeq 0.41 m^4_{\text{Pl}}.  \label{rhoc}
\end{equation}  

In addition, $\mathbf{\Omega}$ becomes negative-valued, leading to an effective change of signature of the metric
(Euclidean phase) around the bounce. The interested reader will find a technical discussion in \cite{martin_deformed} and some qualitative speculations in \cite{jakub_bkl}. Intuitively, this signature change can be straightforwardly interpreted as a change of sign of the Poisson bracket between Hamiltonian constraints.
Equation (\ref{eomk}) is apparently ill-defined as ${\bf\Omega}'/{\bf\Omega}\to\infty$ at $\eta=\eta^{(-)}$ and 
$\eta=\eta^{(+)}$, the values of conformal time when $\rho=\rho_c/2$ before and after the bounce, respectively. 
However, regular solutions do exist by rewriting Eq. (\ref{eomk}) as:

\begin{equation}
{h}'=\mathbf{\Omega} g~;~{g}'=-2\Hc g-k^2h,
\label{eq1}
\end{equation}

which is regular.

\begin{figure}
	\begin{center}
		\includegraphics[scale=1]{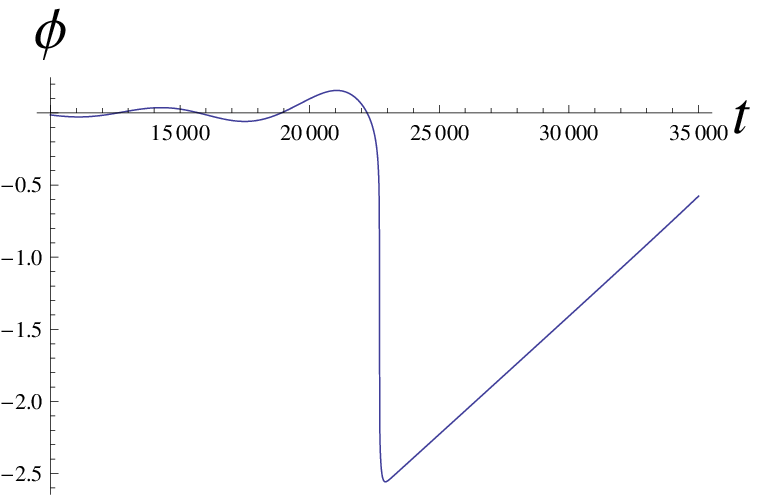} \\
		\includegraphics[scale=1]{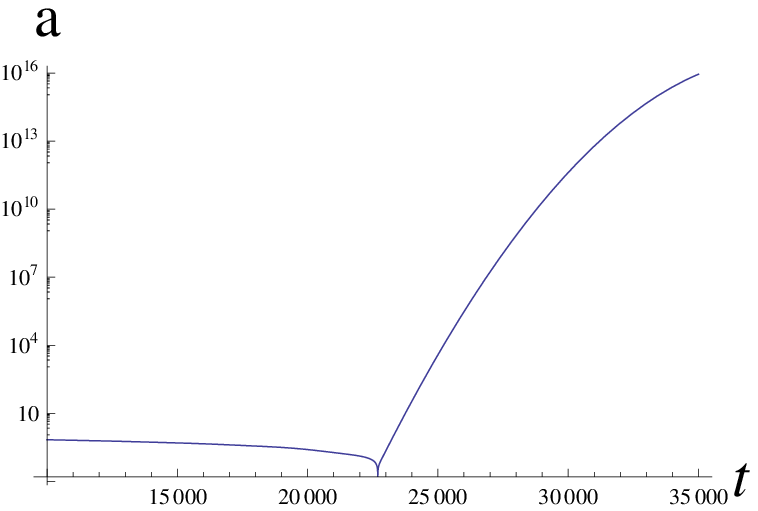}
		\caption{Evolution of the scalar field (upper panel) and the scale factor (lower panel) as a 
		function of cosmic time (the bounce corresponds to $t= 22 693$). The parameters are 
		$m=10^{-3}M_{\mathrm{Pl}}$ and $x_B=-1.5\times10^{-3}$.}
		\label{fig2}
			\end{center}
\end{figure}

The same set of equations in cosmic time is:
\begin{equation}
\dot{h}=\frac{\mathbf{\Omega}}{a} g~;~\dot{g}=-2Hg-\frac{k^2}{a}h,
\label{eq1cosmic}
\end{equation}

where $dot{}$ means differentiation with respect to cosmic time and $H$ is the usual Hubble parameter.
The dynamics can also be recast in a single second-order equation:

\begin{equation}
{g}''+2\Hc{g}'+(2{\Hc}'+\mathbf{\Omega} k^2)g=0.
\label{eq2}
\end{equation}

Whatever the chosen form, either Eq. (\ref{eq1}), (\ref{eq1cosmic}) or  (\ref{eq2}), the evolution can be computed 
numerically. Of course, the propagation of modes has to be coupled with the background evolution which is 
drastically modified by the holonomy corrections that are at the origin of the bounce. The cosmological background 
evolution is basically driven by a single scalar massive matter field of mass $m$. We define

\begin{equation}
x := \frac{m\phi}{\sqrt{2\rho_{\text{c}}}} \  \text{and} \  y :=\frac{\dot{\phi}}{\sqrt{2\rho_{\text{c}}}}, \label{xy}
\end{equation}

which, respectively, represent the density of potential and kinetic energy normalized so that $x_B^2+y_B^2 = 1$ at the 
bounce. The free parameters of the study are, therefore, $m$, $x_B$ (the value of $x$ at the bouce) 
and the relative sign of $x_B$ and $y_B$. 
Interestingly, if the initial conditions for the background are specified at any time, long enough 
before the bounce, the probability of $|x_B|$ is strongly peaked around a given value of order $m$ 
(in Planck units), with $\mathrm{sign}(x_B)=\mathrm{sign}(y_B)$ (the detailed probability
distribution for $x_B$ will be 
studied somewhere else \cite{linda}). For numerical reasons, it is better to specify
computational initial conditions for the background before the bounce rather than 
at the bounce. Because of the peaked probability, the resulting $x_B$ is always close to the 
same value.

It is also necessary to assign a numerical value to the scale factor, 
$a$ at some point. This
choice has, of course, no physical consequences but has to be taken into account for
the interpretation of the meaning of  the wave vectors $k$, since they are expressed in the coordinate space
and not in the physical space. The explicit choice made was $a=1$ at the bounce, which
is numerically easier than the usual normalization at the nowadays value.\\

In  Fig. \ref{fig2}, the evolution of the scalar field and scale factor 
are shown for some typical parameters. As expected, the oscillations of the scalar field are amplified before the 
bounce, because the negative Hubble parameter acts as an antifriction term. Then, just after the bounce, 
the Hubble parameter becomes positive and large, acting as a huge friction and, therefore, leading to slow roll 
inflation.

\begin{figure}
	\begin{center}
		\includegraphics[scale=1]{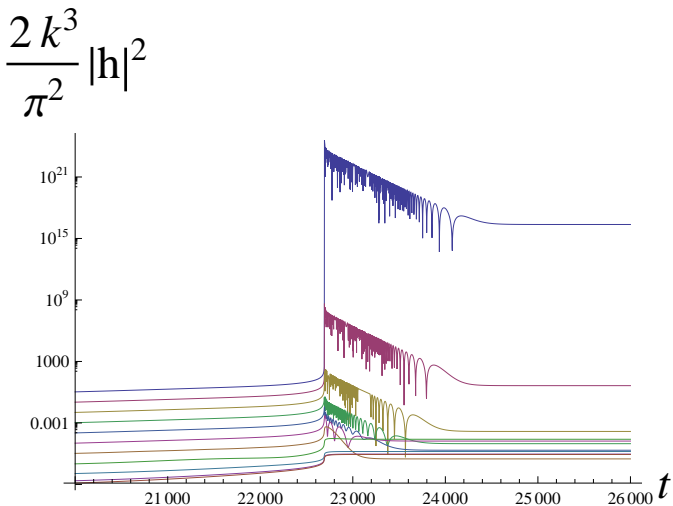}
		\caption{Mode amplitudes as a function of time, corresponding (from top to bottom at $t=20 000$) to
		$k=10^{2},10^{1.5},10^{1},10^{0.5},1,10^{-0.5},10^{-1},10^{-1.5},10^{-2},10^{-2.5}, and 10^{-3}$.
		The parameters are 
		$m=10^{-3}M_{\mathrm{Pl}}$ and $x_B=-1.5\times10^{-3}$. It should be noticed that the 
		initial conditions fore each mode are specified long before the time interval of 
		this plot.}
		\label{fig4}
	\end{center}
\end{figure}

\begin{figure*}
	\begin{center}
		\includegraphics[scale=1]{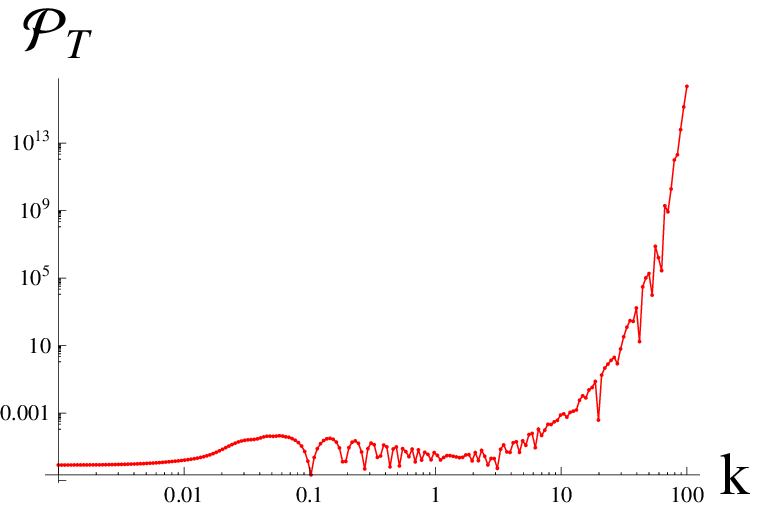} \includegraphics[scale=1]{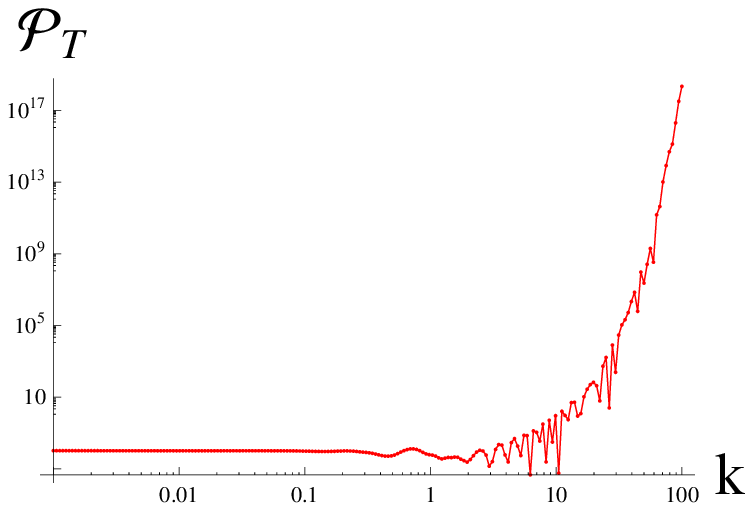}
		\caption{Power spectrum of tensor perturbations after inflation for $m=10^{-3}M_{\mathrm{Pl}}$
		(left panel), and for $m=10^{-1}M_{\mathrm{Pl}}$ (right panel).}
		\label{fig5}
	\end{center}
\end{figure*}

The amplitudes of some Fourier modes of $h$ are plotted in Fig. \ref{fig4}. They are obtained by choosing the Minkowski vacuum as the initial state, since $z''/z\to 0$ in the remote past.

Before the bounce, for 
$k^2 \gg z''/z$, $|h|^2=1/(2ka^2)$, when $z''/z\approx k^2$ or $z''/z > k^2$, $|h|^2$ grows more quikly.
Since the amplitudes of smaller $k$ start growing more quickly before the amplitudes of larger $k$, 
this adds up to a collecting effect that brings all modes up to a certain 
$k\approx \max_{t<t_B}(\sqrt{z''/z})$ up to the same amplitude. After the bounce, the amplitudes 
oscillate until $k^2 \gg z''/z$ when we get $v\propto a$ (as can bee seen from Eq. (\ref{EoM})) and, 
therefore, $h=\text{constant}$.

Finally, the power spectra for different cases are presented in Fig. \ref{fig5}. The main features are the following:

\begin{itemize}
\item a flat (scale invariant) infrared limit, 
\item an oscillating intermediary part,
\item an exponential behavior in the ultraviolet limit (starting around $k=2$ independently of $m$).
\end{itemize}

This obviously exhibits important deviations, with respect both to the standard GR case and with respect to previous 
LQC computations without 
the $\mathbf{\Omega}$ term. Although surprising at first sight, the exponential divergence in the UV limit might not be 
catastrophic as physics at a very small scale is anyway not described by the primordial power spectrum.

Furthermore, this ultraviolet behavior can be checked analytically. In the large $k$ limit of Eq. (\ref{EoM}), the 
WKB conditions are satisfied in the Euclidean phase around the bounce. More precisely, those WKB conditions are met 
for $\eta\in[\eta^{(-)}+\epsilon^{(-)}_k,\eta^{(+)}-\epsilon^{(+)}_k]$ with 
$\epsilon^{(\pm)}_k\sim(k^2\left|\Omega'(\eta=\eta^{(\pm)}\right|)^{-1/3}$. The Mukhanov-Sasaki function can be 
approximated by 

\begin{equation}
v_\mathrm{T}=v_+e^{ik\int\sqrt{\mathbf{\Omega}}d\eta}+v_-e^{-ik\int\sqrt{\mathbf{\Omega}}d\eta}.
\end{equation}

As $\bf\Omega$ is negative-valued during the Euclidean phase, the tensor mode is dominated by its 
exponentially growing solution

\begin{equation}
	h\propto\exp\left(k\int_{\eta^{(-)}+\epsilon^{(-)}_k}^{\eta^{(+)}-\epsilon^{(+)}_k}\sqrt{\left|{\bf\Omega}\right|}d\eta\right).
\end{equation}

This can also be seen in Fig. \ref{fig4} where the amplitude of large $k$ modes 
grows rapidly in the vicinity of the bounce, where $\mathbf{\Omega}<0$.

%In the large $k$

%limit the solutions of Eq. (\ref{eqh}) are indeed:

%\begin{equation}

%h=h_+e^{ik\int\sqrt{\mathbf{\Omega}}}+h_-e^{-ik\int\sqrt{\mathbf{\Omega}}} 

%\propto \left(e^{\int \mathrm{Im}\left(\sqrt{\mathbf{\Omega}}\right)}\right)^k.

%\end{equation}

%This can also be seen in Fig. \ref{fig4} where the amplitude of large $k$ modes 

%grows rapidly in the vicinity of the bounce, where $\mathbf{\Omega}$ is negative.

%\begin{figure}[ht]

%	\begin{center}

%		\includegraphics[scale=0.7]{spectrum1.pdf}

%		\includegraphics[scale=0.7]{spectrum1.eps}

%		\includegraphics[scale=0.45]{speclin.pdf}	

%		\caption{Spectrum of tensor perturbation amplidues after inflation for $m=10^{-3}$ and 

%		$x_B=5\times 10^{-4}$.}

%		\label{fig5}

%	\end{center}

%\end{figure}

%\begin{figure}[ht]

%	\begin{center}

%		\includegraphics[scale=0.7]{spectrum2.pdf}

%		\includegraphics[scale=0.7]{spectrum2.eps}

%		\includegraphics[scale=0.45]{speclin.pdf}	

%		\caption{Spectrum of tensor perturbations for $m=10^{-3}$ and $x_B=5\times 10^{-3}$.}

%		\label{fig6}

%	\end{center}

%\end{figure}

\section{Dicussion}

This study implements in a consistent way the modified algebra induced by holonomy corrections in the calculation of the 
primordial tensor power spectrum. Thanks to numerical calculations, it was possible to solve the equation of motion for 
gravitational waves. The resulting spectrum exhibits specific features. Of course, this raises important questions. 
First, the well-known problem of trans-Planckian modes in inflation (see, {\it e.g.}, \cite{trans}) should be treated with a 
specific care in LQG in which the very meaning of a length smaller than the Planck length is dubious. If the number of 
e-folds of inflation is chosen (by appropriately setting a very small fraction of potential energy density at the bounce)
to be just above the minimum required value, then modes relevant for the CMB are still sub-Planckian and the approach 
makes sense anyway. In other cases, the effective theory might just break down. With the normalization chosen in this work
the trans-Planckian window corresponds to $k>1$. Second, the propagation of modes through
the Euclidean phase is not straitghforward \cite{martin_deformed}. Strictly speaking, there is no "time" in that region 
and the concept of evolution is not well-defined. In this work, we have deliberately chosen to withdraw the conceptual
issues associated with the transition between hyperbolic and elliptic solutions and to focus on a well defined 
mathematical solution. An alternative approach, based on the BKL conjecture, will be studied later \cite{jakub}.
An analogous study should also be performed for scalar modes. The regularization trick used here, however, does not 
apply directly, and other methods have to be constructed. We stress othat the case of scalar modes with 
holonomy corrections has been studied in \cite{yue} and \cite{wilson-ewing} but in different settings for the 
background ; for the study of \cite{yue} is restricted to superinflation while the study 
of \cite{wilson-ewing} considered a dustlike bouncing Universe. Finally, those results will have to be 
compared with forthcoming studies based on other very recent approaches to LQC \cite{news}.

\section*{Ackowledgments}

We would like to thank Arnaud Demion with whom the first ideas were discussed.
T.C. was supported by the NSF grant PHY-1205388. This work was supported by the Labex ENIGMASS.

\end{document}